\documentclass{ws-procs975x65}

\begin{document}

% to switch ON running title
% \markboth{Yu. V. Dumin}{Testing the Dark-Energy-Dominated Cosmology
% by the Solar-System Experiments}

% \wstoc{Testing the Dark-Energy-Dominated Cosmology
% by the Solar-System Experiments}{Yu. V. Dumin}

\title{TESTING THE DARK-ENERGY-DOMINATED COSMOLOGY
BY THE SOLAR-SYSTEM EXPERIMENTS}

\author{YURII V. DUMIN}

\address{Theoretical Department, IZMIRAN, Russian Academy of Sciences,
Troitsk, 142190 Russia\\
\email{dumin@yahoo.com}}

\bodymatter

According to the recent astronomical data, the most part of energy
in the Universe is in the `dark' form, which is effectively described
by $\Lambda$-term in Einstein equations. All arguments in favor
of the dark energy were obtained so far from the observational data
related to very large (intergalactic) scales. Is it possible to find
a manifestation of the dark energy at much less scales ({\it e.g.}
inside the Solar system)?

In general, such effects can be expected from the solution of
the equations of General Relativity (GR) for a point-like mass~$M$
in the $\Lambda$-dominated (de~Sitter) Universe, which was obtained
by Kottler\cite{kot18} very long time ago. The presence of
$\Lambda$-term should change, particularly, the standard relativistic
shift of Mercury's perihelion. This was the idea by Cardona \&
Tejeiro\cite{car98}, who proposed using the measure of the uncertainty
in our knowledge of Mercury's perihelion shift to impose the upper
bound on~$\Lambda$. The result obtained was not so good as other
cosmological estimates but, surprisingly, the accuracy was worse by
only $1{\div}2$~orders of magnitude. A more skeptical viewpoint on
the same subject was presented recently by Iorio\cite{ior06}.

Since accuracy of the above method is insufficient, it was proposed
in our previous papers \cite{dum01,dum03} to utilize the data of radial
(rather than angular) measurements of the Moon to reveal the anomalous
increase in its distance from the Earth produced by the $\Lambda$-term,
which looks formally as `local' Hubble expansion. Why is it necessary
to reexamine the problem of local Hubble expansion just in the context
of `dark-energy'-dominated cosmological models?

Hubble dynamics at small scales is studied for a long time, starting
from the pioneering work by McVittie\cite{mvi33}. Although the results
by various authors were quite contradictory ({\it e.g.} review by
Bonnor\cite{bon00} and references therein), the most popular point
of view was that the Hubble expansion manifests itself only at
the sufficiently large distances (from a few Mpc) and is absent at
the less scales\cite{mis73}. There were a few arguments in favor of such
conclusion, such as the so-called Einstein--Straus theorem\cite{ein45},
a quasi-Newtonian treatment of Hubble effect in a small volume as
a tidal-like action by distant matter ({\it e.g.} the recent work by
Dom\'{i}nguez \& Gaite\cite{dom01} and references therein), and
the Einstein--Infeld--Hoffmann (EIH) surface integral method,
which was applied to the problem of local Hubble expansion by
Anderson\cite{and95}. Unfortunately, as is shown in
Ref.~\refcite{dum05}, all these approaches become inapplicable when
the Universe evolution is governed by $\Lambda$-term, uniformly
distributed in space.

A frequent experimental argument against the Hubble expansion within
Solar system is based on the available constraint on time variation
in the gravitational constant derived from the lunar dynamics, which
is now as strong as
$ \, {\dot G} / G = \! (4{\pm}9){\times}10^{-13} $~yr${}^{-1}$
(Ref.~\refcite{wil04}). Unfortunately, the equivalence between
the effect of variable~$G$ and the cosmological expansion, stated
by some authors, is based solely on the Newtonian arguments. A more
accurate treatment of this problem in the GR framework\cite{dum07}
shows that manifestation of $\Lambda$-term in some components of
the metric tensor really looks like the influence of variable~$G$
if we assume that $ G = G_0 + {\dot G} \, t $, where
$ {\dot G} \! = \! - c \sqrt{\Lambda / 3} \: $; but such interpretation
is not self-consistent: the $\Lambda$-dependence of a few other
components is not expressible in terms of the variable coefficient of
gravitational coupling. Therefore, the available limits on
$ \, {\dot G} / G \, $, in general, cannot be reinterpreted as
a constraint on local cosmological dynamics.

Since all the commonly-used arguments against the local Hubble
expansion fail in the case of dark energy, it becomes reasonable
to seek for the corresponding effect; and the most sensitive tool
seems to be the lunar laser ranging (LLR).\cite{dic94,nor99}
For example, if we assume that planetary systems experience
Hubble expansion with the same rate as everywhere in the Universe
($60{\div}70$~km{\,}s${}^{-1}${\,}Mpc${}^{-1}$), then average radius
of the lunar orbit $R \,$ should increase by $\sim \! 50$~cm for
the period of 20~years. On the other hand, the accuracy of LLR
during the last 20~years was not worse than $2{\div}3$~cm; so
the perspective of revealing the local Hubble effect looks very good.

The main problem is to exclude the effect of geophysical tides, which
also contributes to the secular increase in the Earth--Moon distance as
$ \dot R = k \, {\dot T}_{\rm E} \, $,
where $T_{\rm E}$~is the Earth's diurnal period, and
$k \! = \! 1.81{\times}10^5$~cm{\,}s${}^{-1}$ ({\it e.g.}
Ref.~\refcite{dum03}). So, if ${\dot T}_{\rm E}$ is known from
independent astrometric measurements of the Earth's rotation
deceleration with respect to distant objects, then the above relation
can be used to exclude the geophysical tides and, thereby, to reveal
a probable Hubble expansion.

The telescopic data, accumulated from the middle of the 17th~century,
were processed by a few researches; and one of the most detailed
compilations was presented recently in Ref.~\refcite{sid02}.
Of course, the value of secular trend derived from the quite short
time series can suffer from considerable periodic and quasi-periodic
variations in $T_{\rm E}$. So, the main aim of our statistical analysis,
described in more detail in Ref.~\refcite{dum05}, was to estimate as
carefully as possible the `mimic' effect of such influences. The result
can be written as
$ {\dot T}_{\rm E} \! =
  \! (8.77{\pm}1.04){\times}10^{-6}~{\rm s}{\,}{\rm yr}^{-1} \!\! $ .
(This value is appreciably less than in our previous work\cite{dum03},
where it was taken from the older literature.)

\begin{table}[t]
\tbl{Rates of secular increase in the mean Earth--Moon distance.}
{\begin{tabular}{@{}lll@{}}
\hline\\
&&\\[-18pt]
\multicolumn{1}{c}{Method} &
\multicolumn{1}{c}{Immediate measurement by} &
\multicolumn{1}{c}{Independent estimate from the} \\
& \multicolumn{1}{c}{the lunar laser ranging} &
\multicolumn{1}{c}{Earth's tidal deceleration}
\\[1pt]
\hline\\
&&\\[-18pt]
\hphantom{i} Effects involved &
(1) geophysical tides &
(1) geophysical tides \\
& (2) local Hubble expansion & \\
&&\\[-7pt]
\hphantom{i} Numerical value &
$3.8{\pm}0.1$~cm{\,}yr${}^{-1}$ &
$1.6{\pm}0.2$~cm{\,}yr${}^{-1}$ \\[1pt]
\hline
\end{tabular}}
\label{tab:Compar}
\end{table}

The entire analysis of LLR vs.\ the astrometric data is summarized in
Table~\ref{tab:Compar}. The excessive rate of increase in the lunar
orbit, $2.2{\pm}0.3$~cm{\,}yr${}^{-1}$, can be attributed just to
the local Hubble expansion with rate
$ H_0^{\rm (loc)} \! = 56{\pm}8~{\rm km{\,}s^{-1}{\,}Mpc^{-1}} \! $ .

Next, it is reasonable to assume that the local Hubble expansion is
formed only by the uniformly-distributed dark energy, while the
irregularly-distributed (aggregated) forms of matter begin to affect
the Hubble flow at the larger distances, thereby increasing its rate
up to the standard intergalactic value. If the Universe is spatially
flat and filled with the $\Lambda$-term and a dust-like (`cold')
matter, with densities $\rho_{{\Lambda}0}$ and $\rho_{{\rm D}0}$
respectively, then\cite{lan75}
\begin{equation}
H_0 = \, \sqrt{\frac{8 \pi G}{3}}
      \; \sqrt{\, \rho_{{\Lambda}0} + \rho_{{\rm D}0}} \; .
\label{eq:H_0-rho}
\end{equation}
So, if $H_0$ is formed locally only by ${\rho}_{{\Lambda}0}$,
while globally by both these terms, ${\rho}_{{\Lambda}0}$ and
${\rho}_{{\rm D}0}$ (or, in terms of the relative densities,
$ {\Omega}_{{\Lambda}0} = {\rho}_{{\Lambda}0} / {\rho}_{\rm cr} $
and $ {\Omega}_{{\rm D}0} = {\rho}_{{\rm D}0} / {\rho}_{\rm cr} $),
then
\begin{equation}
\frac{H_0^{\rm (loc)}}{H_0} =
{\left[ 1 +
  \frac{\Omega_{{\rm D}0}}{\Omega_{{\Lambda}0}} \, \right]}^{-1/2} \! .
\end{equation}

At $\Omega_{{\Lambda}0} \! = \! 0.75$ and
$\Omega_{{\rm D}0} \! = \! 0.25$, we get
${H_0} / {H_0^{\rm (loc)}} \! \approx 1.15$. Therefore,
$ H_0 = 65{\pm}9~{\rm km{\,}s^{-1}{\,}Mpc^{-1}} \!\! $ ,
which is in reasonable
agreement both with the well-known WMAP result,
$71{\pm}3.5$~km{\,}s${}^{-1}${\,}Mpc${}^{-1}$, and with the recent
Hubble diagram for type Ia supernovae\cite{rei05}, whose
interpretation requires a slightly reduced value of $H_0$.

Therefore, the presence of local Hubble expansion, caused by the
$\Lambda$-term, gives us a reasonable explanation of the anomalous
increase in the lunar orbit, consistent with the `large-scale'
astronomical data. Thereby, this is one more argument in favor of
the dark energy. Besides, if the local Hubble expansion really exists,
it should result in profound consequences not only for cosmological
evolution but also for the dynamics of planetary systems and other
`small-scale' astronomical phenomena.

% \bibliographystyle{ws-procs975x65}

% \bibliography{Dumin}

\begin{thebibliography}{10}

\bibitem{kot18}
F.~Kottler, {\em Ann.\ Phys.} {\bf 56}, p. 401 (1918).

\bibitem{car98}
J.~F. Cardona and J.~M. Tejeiro, {\em Astrophys.\ J.} {\bf 493}, p.~52 (1998).

\bibitem{ior06}
L.~Iorio, {\em Int.\ J.\ Mod.\ Phys.\ D} {\bf 15}, p. 473 (2006).

\bibitem{dum01}
Yu.~V. Dumin, {\em Geophys.\ Res.\ Abstr.} {\bf 3}, p. 1965 (2001).

\bibitem{dum03}
Yu.~V. Dumin, {\em Adv.\ Space Res.} {\bf 31}, p. 2461 (2003).

\bibitem{mvi33}
G.~C. McVittie, {\em MNRAS} {\bf 93}, p. 325 (1933).

\bibitem{bon00}
W.~B. Bonnor, {\em Gen.\ Rel.\ Grav.} {\bf 32}, p. 1005 (2000).

\bibitem{mis73}
C.~W. Misner, K.~S. Thorne and J.~A. Wheeler, {\em Gravitation}
  (W.H.~Freeman~\& Co., San Francisco, 1973).

\bibitem{ein45}
A.~Einstein and E.~G. Straus, {\em Rev.\ Mod.\ Phys.} {\bf 17}, p. 120 (1945).

\bibitem{dom01}
A.~Dom\'{i}nguez and J.~Gaite, {\em Europhys.\ Lett.} {\bf 55}, p. 458 (2001).

\bibitem{and95}
J.~L. Anderson, {\em Phys.\ Rev.\ Lett.} {\bf 75}, p. 3602 (1995).

\bibitem{dum05}
Yu.~V. Dumin, astro-ph/0507381.

\bibitem{wil04}
J.~G. Williams, S.~G. Turyshev and D.~H. Boggs, {\em Phys.\ Rev.\ Lett.} {\bf
  93}, p. 261101 (2004).

\bibitem{dum07}
Yu.~V. Dumin, {\em Phys.\ Rev.\ Lett.} {\bf 98}, p. 059001 (2007).

\bibitem{dic94}
J.~O. Dickey, {\em et al.}, {\em Science} {\bf 265}, p. 482 (1994).

\bibitem{nor99}
K.~Nordtvedt, {\em Class.\ Quant.\ Grav.} {\bf 16}, p. A101 (1999).

\bibitem{sid02}
N.~S. Sidorenkov, {\em Physics of the Earth's Rotation Instabilities}
  (Nauka-Fizmatlit, Moscow, 2002, in Russian).

\bibitem{lan75}
L.~D. Landau and E.~M. Lifshitz, {\em The Classical Theory of Fields}
  (Pergamon, Oxford, 1975).

\bibitem{rei05}
B.~Reindl, G.~A. Tammann, A.~Sandage and A.~Saha, {\em Astrophys.\ J.} {\bf
  624}, p. 532 (2005).

\end{thebibliography}

% \vfill

\end{document}